\documentclass[aps,prd,superscriptaddress,nofootinbib]{revtex4-2}

\usepackage{supertabular} 
\usepackage{placeins}
\usepackage{epsfig}
\usepackage{graphicx}
\usepackage{tensor}

\usepackage{float}
\usepackage{mathtools}
\usepackage{array}
\usepackage{dcolumn}

\usepackage{latexsym}
\usepackage{amsmath}
\usepackage{amssymb}
\usepackage{amsfonts}
\usepackage{bm}
\usepackage{physics}

\usepackage{color}
\definecolor{purple}{rgb}{0.5,0,0.5}
\definecolor{blue}{rgb}{0.0,0,0.9}
\definecolor{prdblue}{rgb}{0.133,0.118,0.498}
\usepackage[colorlinks=true, pdfstartview=FitV, linkcolor=prdblue, citecolor= prdblue, urlcolor=prdblue]{hyperref}

\allowdisplaybreaks
\begin{document}


\title{Gross–Pitaevskii–Poisson equations with a $\xi R \phi^4$ non-minimal coupling term}

\author{Bryan Cordero-Patino}
\email[]{bcorpat@alu.upo.es}
\affiliation{Departamento de Sistemas F\'isicos, Qu\'imicos y Naturales, Universidad Pablo de Olavide, E-41013 Sevilla, Spain}

\author{\'Alvaro Duenas-Vidal}
\email[]{alvaro.duenas@epn.edu.ec}
\affiliation{Departamento de Física, Escuela Politécnica Nacional, Quito 170143, Ecuador}

\author{Jorge Segovia}
\email[]{jsegovia@upo.es}
\affiliation{Departamento de Sistemas F\'isicos, Qu\'imicos y Naturales, Universidad Pablo de Olavide, E-41013 Sevilla, Spain}

\date{\today}

\begin{abstract}
In scenarios where the Peccei–Quinn symmetry breaks after inflation, small-scale axion inhomogeneities may gravitationally collapse into bound structures. The evolution of these systems is typically modeled through cosmological perturbation theory applied to the Einstein–Klein–Gordon equations. In the non-relativistic regime, this framework reduces to the Gross–Pitaevskii–Poisson or Schrödinger–Poisson equations, depending on whether axion self-interactions are taken into account. In this work, a non-minimal gravitational coupling term $\xi R \phi^4$ is included into the axion's relativistic action as a way to introduce a gravitationally mediated pairwise interaction. By performing a perturbative expansion and subsequently taking the non-relativistic limit, an alternative set of equations that govern the early stages of structure formation is obtained.
\end{abstract}


\maketitle


\section{INTRODUCTION}
\label{sec:introduction}

Dark matter (DM) remains as one of the clearest indications of physics beyond the Standard Model (SM), shaping developments across particle theory, astrophysics, and cosmology. The axion, originally just a byproduct of the Peccei–Quinn solution to the strong CP problem in Quantum Chromodynamics (QCD) \cite{Peccei::1977,Peccei_2::1977}, has gained remarkable attention as a DM candidate in recent years in both experimental and theoretical studies \cite{Battat::2024,Jackson::2023,ParticleDataGroup::2022}. At energy scales below which QCD non-perturbative effects become significant, the axion can be effectively described by a real scalar field $\phi$ under the Klein–Gordon action with a non-trivial interaction potential $V(\phi)$,

\begin{equation}
    S = - \int d^4 x \, \left( \frac{1}{2} \partial_\mu \phi \partial^\mu \phi + V(\phi) \right) .
    \label{eq::axion-rltvs_action}
\end{equation}

\noindent
The axion potential can be derived from chiral perturbation theory \cite{Luzio::2020}. Since $V(\phi)$ is an even function of $\phi$, it can be expanded in powers of $\phi^2$ around $0$, with the quadratic term fixing the axion mass $m_a$. It can be shown that $m_a = \sqrt{\chi}/f_a$, where $\chi$ is the topological susceptibility of QCD and $f_a$ is the axion decay constant. The most precise computation to date yields \cite{ParticleDataGroup::2022,snowmass_axions::2022}

\begin{equation}
    m_a = 5.691(51) \left(\frac{10^{12} \text{GeV}}{f_a} \right) \mu \text{eV} \, .
\end{equation}

\noindent
The viable DM mass window for the axion is $m_a \in (10^{-6} - 10^{-3}) \, \text{eV}$ \cite{Irastorza::2021}. The higher-order terms of $V(\phi)$ encode the axion self-interaction couplings. Since $m_a$ and $f_a$ dictate most of the axion's properties and interactions, it is practical to express the higher-order terms using these parameters,

\begin{equation}
    V(\phi) = \frac{1}{2} m_a^2 \phi^2 + \left(m_a f_a \right)^2 \sum_{n=2}^{\infty} \frac{\lambda_{2n}}{(2n)!} \left(\frac{\phi}{f_a} \right)^{2n} \, ,
    \label{eq::axion-rltvs_potential}
\end{equation}

\noindent
where $\lambda_{2n}$ are dimensionless coupling constants of $\mathcal{O}(1)$. The Z$_2$ symmetry of $V(\phi)$ ensures that the number of axions in a scattering reaction is conserved modulo 2. Among these couplings, the axion-axion interactions are predominant within the pure axion sector. Notably, these pairwise interactions are attractive. The parametrization in Eq.~(\ref{eq::axion-rltvs_potential}) indicates that $m_a^2/f_a^2$ acts as a quantum-loop factor. Within the axion DM mass window, this factor lies in $m_a^2/f_a^2 \in (10^{-52} - 10^{-44})$. Such an exceedingly small value emphasizes the negligible effect of quantum-loop corrections, therefore validating the use of classical field theory for the axion description.  \cite{Braaten::2019}. 

While most axion detection efforts focus on direct detection experiments, several indirect search strategies have been proposed. These approaches seek to detect astrophysical and cosmological effects arising from the axion field presence. In particular, if axion inhomogeneities survive the inflationary period, localized bound structures can emerge from them. Axion miniclusters, also known as axion minihalos, are gravitationally bound and virialized objects decoupled from the Universe's expansion that originate from density fluctuations around the time of matter–radiation equality \cite{Braaten::2019,Chang::2024}. Axion stars, by contrast, are stable bound states where the gravitational attraction is balanced by gradient pressure. These configurations are categorized into two branches: a dilute branch, where stability involves only gravity and kinetic pressure, and a dense branch, where self-interactions become significant beyond a critical mass threshold. In addition, several authors have suggested the existence of self-bound configurations known as axitons \cite{Chang::2024,Zhang::2018,Braaten::2019}. 

DM halo regions hosting axion miniclusters may reach mass densities several orders above the halo average. Such pronounced densities could give rise to detectable gravitational lensing effects, offering an avenue to constrain their abundance. Nevertheless, further simulations are required to assess the feasibility of this strategy. An alternative detection strategy relies on magnetometer networks sensitive enough to detect the passage of compact DM objects through the Earth's vicinity. Further observable signatures may result from head-on collisions of axion stars with other astrophysical objects  \cite{Braaten::2019,Jackson_Kimball::2018,ParticleDataGroup::2022}. Recently, the James Webb Space Telescope has reported observations supporting the existence of dark stars, a class of early stellar objects composed of hydrogen and helium clouds whose gravitational collapse is delayed by DM heating mechanisms. They are typically modeled using WIMPs or self-interacting DM \cite{Ilie::2025,Wu::2022}. An analogous concept, termed a Hydrogen axion star, in which dense axion stars serve as the heat source, was proposed in Ref.~\cite{Bai::2016}. Consequently, modeling the evolution and the observable implications of compact axion configurations continues being a promising DM research avenue.

Since DM is mostly non-relativistic, the axion cosmological role is best described by a non-relativistic effective field theory (NREFT). Several methods have been developed to construct this low energy limit. Herein, an approach involving an explicit field redefinition that maps the real scalar field to a complex scalar field is adopted \cite{Namjoo::2017,Salehian::2020,Salehian::2021,Cordero::2023}. Moreover, an accurate model of axion gravitational collapse must rely on General Relativity (GR). While the unification of Einstein's theory of gravity and particle physics remains unsolved, curvature effects are often incorporated via the minimal coupling prescription. This approach leads to the Einstein-Klein-Gordon equations that are widely used to characterize macroscopic axion structures. Notably,
cosmological perturbation theory is used to study their formation, with the low-energy limit subsequently reached through the aforementioned field redefinition. This results in the Gross–Pitaevskii–Poisson system when self-interactions are present, and the Schrödinger–Poisson equations when they are neglected \cite{Jackson::2023}.

Due to the inherently attractive nature of axion pairwise interactions, gravitationally bound configurations cannot grow as extensively as they might if the interactions were repulsive. Large-scale objects require some form of repulsive force or pressure to counteract gravitational collapse. Heisenberg's uncertainty principle provides this support for bosons, although its stabilization impact is weaker than the one provided by Pauli's exclusion principle in neutron stars. Consequently, the maximum mass of bosonic stars with attractive self-interactions is relatively low \cite{Chavanis::2011,Eby::2016}. In this work, a non-minimal coupling term proportional to $R\phi^4$, where $R$ is the scalar of curvature, is introduced as it naturally encodes gravitationally mediated pairwise interactions. Furthermore, the inclusion of the $R\phi^4$ term may facilitate the formation of more complex axion structures. Since $R$ varies with the local geometry, the non-minimal coupling term can produce either attractive or repulsive effects, potentially enhancing both the stability and mass range of axion stars. After establishing the non-relativistic equations relevant to the early-time behavior of structure formation, the Hamiltonian that governs a system of $N$ axions is obtained. This provides a foundation for analyzing the thermodynamics of the axion field during gravitational collapse, as the corresponding partition function can now be computed.


\section{GRAVITY COUPLING}
\label{sec:gravity_coupling}

The simplest way to construct an action that remains invariant under general coordinate transformations is to begin with the flat-spacetime action and to apply the minimal coupling prescription \cite{Leonard::2009,Padmanabhan::2010}:

\begin{itemize}
    \item Replace all ordinary derivatives $\partial_\mu$ with covariant derivatives $\nabla_\mu$.
    \item Substitute the Minkowski metric $\tensor{\eta}{_\mu_\nu}$ with the general metric $\tensor{g}{_\mu_\nu}$.
    \item Swap $d^4 x$ with the invariant volume element $d^4 x \sqrt{-g}$, where $g = \det (\tensor{g}{_\mu_\nu})$.
\end{itemize}

\noindent
Applying this procedure to the axion Lagrangian in Eq.~(\ref{eq::axion-rltvs_action}) yields

\begin{equation}
    S_m = - \int d^4 x \, \sqrt{-g} \left(\frac{1}{2} \, \nabla^\mu \phi \nabla_\mu \phi + V(\phi) \right) \,.
    \label{eq::action_m_1}
\end{equation}

\noindent
The subscript $m$ indicates that this action corresponds to the matter sector of the system. Meanwhile, the dynamics of the gravitational field are described by the Einstein–Hilbert action\footnote{The broadest form of this action includes $(R-\Lambda)$ instead of $R$, where $\Lambda$ is the cosmological constant; however, this term is not considered in the present work.} :

\begin{equation}
    S_g = \frac{1}{16\pi G} \int d^4x \sqrt{-g} \, R \,,
    \label{eq::Einst_Hilbert_action}
\end{equation}

\noindent
where $\tensor{R}{_\mu_\nu}$ is the Ricci tensor and $G$ is the gravitational constant. The total action, $S_{tot} = S_m + S_g$, upon variation with respect to $\delta \tensor{g}{^\mu^\nu}$, yields the Einstein’s field equations\footnote{Additional boundary terms arise in the calculation. While they are usually neglected, in settings where spacetime boundaries are physically relevant, e.g., black hole thermodynamics and the Hamiltonian formulation of general relativity, they must be explicitly included and treated with care.} :

\begin{equation}
    \tensor{G}{_\mu_\nu} = 8\pi G \, \tensor{T}{_\mu_\nu} \,,
    \label{eq::Einstein_field_eqs}
\end{equation}

\noindent
where $\tensor{G}{_\mu_\nu} = \tensor{R}{_\mu_\nu} - \frac{1}{2} R \, \tensor{g}{_\mu_\nu}$ is the Einstein tensor and $\tensor{T}{_\mu_\nu} = \partial_\mu \phi \partial_\nu \phi - \tensor{g}{_\mu_\nu} \left( \frac{1}{2} \, \partial^\alpha \phi \partial_\alpha \phi + V(\phi) \right)$ is the stress-energy tensor. On the other hand, varying $S_{tot}$ with respect to $\delta \phi$ results in the covariant generalization of the Klein–Gordon equation,

\begin{equation}
    \nabla^\mu \nabla_\mu \, \phi - \frac{\partial V}{\partial \phi} = 0 \,.
    \label{eq::KG_eq_curved}
\end{equation}

\noindent
In contrast to the special relativistic case, plane wave solutions and Fourier analysis cannot be generally employed. Instead, it has to be solved on a case-by-case basis depending on the specific symmetries of the metric \cite{Leonard::2009}. When $V(\phi)$ includes only the mass term, Eqs.~(\ref{eq::Einstein_field_eqs}) and~(\ref{eq::KG_eq_curved}) are denoted as the Einstein-Klein-Gordon equations.

While minimal coupling provides a first approximation of how curvature affects a physical system, it is not uncommon to go beyond the standard covariant substitutions. These non-minimal couplings arise naturally in various settings, including QFT in curved spacetime, Higgs inflation models, modified gravity theories, and the low-energy effective actions derived from superstring theory \cite{Leonard::2009,Hwang::2000,Sotiriou::2010,Bezrukov::2008}. Any extra modification must be invariant under diffeomorphisms and respect the internal symmetries of the theory. More often than not, the additional terms involve the Riemann tensor $\tensor{R}{_\mu_\nu_\rho_\sigma}$ and its contractions, which ensures that they vanish in the Minkowski spacetime. Since the $\phi^4$ term in $V(\phi)$ characterizes the axion pair self-interactions, it is quite natural to include a term proportional to $R \phi^4$ to incorporate gravitationally mediated pair interactions to the framework. As a result, the total action becomes

\begin{align}
    S_{tot} =& - \int \, d^4 x \, \sqrt{-g} \left(\frac{1}{2} \, \partial^{\mu} \phi \, \partial_{\mu} \phi + V (\phi) + \xi R \phi^4 \right) +\frac{1}{16\pi G} \int d^4x \sqrt{-g} \, R \nonumber \\
    =& - \int \, d^4 x \, \sqrt{-g} \left(\frac{1}{2} \, \partial^{\mu} \phi \, \partial_{\mu} \phi + V (\phi) \right) +\frac{1}{16\pi G} \int d^4x \sqrt{-g} \, f(\phi) \, R \, ,
    \label{eq::action_non_min_coupling}
\end{align}

\noindent
where $\xi$ is a constant and $f(\phi) = 1 - (16\pi G) \, \xi \phi^4$.

This type of non-minimal coupling can be described in two mathematically equivalent frames \cite{Padmanabhan::2010}. In the Jordan frame, $f(\phi)$ is retained explicitly in the action. Alternatively, the canonical Einstein-Hilbert action is recovered in the Einstein frame by performing a conformal transformation of the metric,

\begin{equation}
    {g'}_{\mu\nu} = (16 \pi G) \, f(\phi) \, {g}_{\mu\nu} \,.
    \label{eq::Jordan_to_Einstein}
\end{equation}

\noindent
Cosmological perturbations can be carried out in either frame. Nevertheless, physical observables are more straightforward to interpret in the Jordan frame since the metric is not rescaled. For this reason, Eq.~(\ref{eq::Jordan_to_Einstein}) was not employed in this work.

Requiring that $\delta S_{\text{tot}} = 0$ under arbitrary variations of $\delta g^{\mu\nu}$, while systematically discarding surface terms as in the Einstein–Hilbert action, yields a modified form of Einstein’s field equations \cite{Padmanabhan::2010},

\begin{equation}
    f(\phi) \tensor{G}{_\mu_\nu} - 8\pi G \, \tensor{T}{_\mu_\nu} = \nabla_\mu \nabla_\nu f - \tensor{g}{_\mu_\nu} \nabla^\alpha \nabla_\alpha f \,.
\end{equation}

\noindent
Or equivalently,

\begin{equation}
    \left(\frac{1}{8\pi G} - 2\, \xi \, \phi^4 \right) \tensor{G}{^\mu_\nu}  - \, \tensor{T}{^\mu_\nu} + 2\, \xi \, \tensor{\mathcal{Z}}{^\mu_\nu} = 0 \,,
    \label{eq::Einstein_field_eqs_non_minimal}
\end{equation}

\noindent
where

\begin{gather}
    \tensor{\mathcal{Z}}{_\mu_\nu} = \nabla_\mu \nabla_\nu \, \phi^4 - \tensor{g}{_\mu_\nu} \, \nabla^\alpha \nabla_\alpha \, \phi^4 \,. 
\end{gather}

\noindent
Meanwhile, varying Eq.~(\ref{eq::action_non_min_coupling}) with respect to $\delta \phi$ implies that

\begin{equation}
    \nabla^\mu \nabla_\mu \phi - \frac{\partial V}{\partial \phi} - 4\, \xi R \, \phi^3 = 0 \,.
    \label{eq::KG_eq_curved_non_minimal}
\end{equation}


\section{THE NEWTONIAN GAUGE}
\label{sec:newtonian_gauge}

Cosmological perturbation theory studies the evolution of small deviations from a homogeneous matter distribution within a background metric. It is crucial for characterizing the growth of density perturbations, the temperature anisotropies and polarization patterns of the Cosmic Microwave Background, and the evolution of gravitational waves. While gravitational collapse can be partially described using Newtonian gravity and classical hydrodynamics, e.g., the Jeans instability criterion, such treatments neglect cosmic expansion. Therefore, a fully consistent description of perturbation growth requires GR \cite{Padmanabhan::2010,Gorbunov::2011}. In the early Universe, perturbations in both the energy density $\delta\rho$ and the metric $\delta \tensor{g}{_\mu_\nu}$ are small compared to their background values, i.e., $\delta\rho \ll \rho$ and $|\delta \tensor{g}{_\mu_\nu}| \ll |\tensor{g}{_\mu_\nu}|$. Under these conditions, a linearized theory of perturbations is both accurate and highly effective. However, linear theory ceases to apply once $\delta\rho / \rho \sim 1$, at which point alternative methods, such as large-scale numerical simulations, are required \cite{Gorbunov::2011}.

It is often convenient to express cosmological perturbations in terms of conformal time $\eta$ instead of cosmological time $t$. The relation between these time coordinates is

\begin{equation}
    a(\eta) \, d\eta = dt \, ,
\end{equation}

\noindent
where $a(\eta)$ is the dimensionless scale factor. Hence, the Hubble parameter is

\begin{equation}
    H(\eta) = \frac{1}{a} \frac{d a}{dt} = \frac{1}{a^2} \frac{da}{d\eta} \, .
\end{equation}

\noindent
The Friedmann–Lemaître–Robertson–Walker (FLRW) metric with zero spatial curvature (k=0) describes a spatially flat, homogeneous, and isotropic expanding Universe,

\begin{equation}
    ds^2 = a^2(\eta) \left[- d \eta^2 + \delta_{ij} \, dx^i dx^j \right] = a^2(\eta) \, \tensor{\eta}{_\mu_\nu} \, dx^\mu dx^\nu \,.
\end{equation}

\noindent
A perturbed version of this metric can be written as

\begin{equation}
    ds^2 = \tensor{g}{_\mu_\nu}(x) \, dx^\mu dx^\nu  = a^2(\eta) \left(\tensor{\eta}{_\mu_\nu} + \tensor{h}{_\mu_\nu}(x)\right) dx^\mu dx^\nu \, ,
\end{equation}

\noindent
where $\tensor{h}{_\mu_\nu}(x)$ encodes the perturbations and, therefore, $|\tensor{h}{_\mu_\nu}| \ll 1$. It is useful to use an overbar to denote background quantities, e.g., the background metric $\tensor{\overline{g}}{_\mu_\nu}(\eta) = a^2(\eta) \, \tensor{\eta}{_\mu_\nu}$. Analogously to the metric tensor, the stress energy tensor $\tensor{T}{_\mu_\nu}$ and the Einstein tensor $\tensor{G}{_\mu_\nu}$ are decomposed into a homogeneous background and a perturbative portion,

\begin{align}
    \tensor{T}{_\mu_\nu}(x) = \tensor{\overline{T}}{_\mu_\nu}(\eta) + \delta \tensor{T}{_\mu_\nu}(x) \, , && \tensor{G}{_\mu_\nu}(x) = \tensor{\overline{G}}{_\mu_\nu}(\eta) + \delta \tensor{G}{_\mu_\nu}(x) \, .
\end{align}

\noindent
Under the minimal coupling prescription, $\tensor{\overline{T}}{_\mu_\nu}$ and $\tensor{\overline{G}}{_\mu_\nu}$ satisfy\footnote{The linearized Einstein equations are normally expressed with one covariant and one contravariant index.}

\begin{align}
    \tensor{\overline{G}}{^\mu_\nu} = 8\pi G \, \tensor{\overline{T}}{^\mu_\nu}\,, && \overline{\nabla}_\mu  \tensor{\overline{T}}{^\mu_\nu} = 0 \,.
\end{align}

\noindent
Meanwhile, the perturbations are described at linear order by

\begin{align}
    \delta\tensor{G}{^\mu_\nu} = 8\pi G \, \delta \tensor{T}{^\mu_\nu}\,, && \delta \left( \nabla_\mu  \tensor{T}{^\mu_\nu} \right)= 0 \,.
\end{align}

\noindent
Furthermore, the background and perturbative counterparts of Eq.~(\ref{eq::KG_eq_curved}) must be included. Since the matter content is described by $\phi(x) = \overline{\phi}(\eta) + \delta \phi(x)$, where $|\delta \phi| \ll |\overline{\phi}|$, $\delta \tensor{T}{^\mu_\nu}$ depends linearly on $\delta \phi$ but exhibits no explicit dependence on $\tensor{h}{_\mu_\nu}$. Similarly, $\delta \tensor{G}{^\mu_\nu}$ includes only terms linear to $\tensor{h}{_\mu_\nu}$.

When focusing on the growth of matter perturbations, only two of the ten degrees of freedom in $\tensor{h}{_\mu_\nu}$ remain physically relevant, after accounting for gauge freedom. The perturbed metric is written as

\begin{equation}
    ds^2 = a^2(\eta) \left[-\left(1+2\,\Phi\right) d\eta^2 + \left(1-2\Psi\right) \delta_{ij} \, dx^i dx^j \right] \,,
    \label{eq::Newtonian_gauge}
\end{equation}

\noindent
where $\Phi(x)$ and $\Psi(x)$ are scalars which follow that $|\Phi|,|\Psi| \ll 1$. This line element is known in literature as the Newtonian gauge \cite{Padmanabhan::2010,Gorbunov::2011}. In the absence of anisotropic stress, i.e., when the traceless part of $\delta \tensor{T}{_i_j}$ vanishes, the linearized Einstein equations naturally enforce $\Phi = \Psi$. An interesting scenario to outline is the evolution of spherically symmetric matter perturbations. In this context, it is advantageous to express the spatial components in spherical coordinates; thereby, the Newtonian gauge becomes

\begin{equation}
    ds^2 = a^2(\eta) \left[-\left(1+2\,\Phi\right) d\eta^2 + \left(1-2\Psi\right) \left(dr^2 + r^2 d\theta^2 + r^2 \sin^2(\theta) d\varphi^2 \right) \right] \,.
    \label{eq::Newtonian_gauge_spherical}
\end{equation}

In the non-minimal coupling framework proposed in this work, the linearization must be applied to Eqs.~(\ref{eq::Einstein_field_eqs_non_minimal}) and (\ref{eq::KG_eq_curved_non_minimal}). Unlike $\delta \tensor{T}{^\mu_\nu}$ and $\delta \tensor{G}{^\mu_\nu}$, the $\xi$-dependent perturbations are linear to both $\delta \phi$ and $\tensor{h}{_\mu_\nu}$. In standard linear perturbation theory, a direct computation of $\tensor{G}{^\mu_\nu}$ can be done by retaining only the first-order perturbative terms through the calculation. This procedure involves using the background metric $\tensor{\overline{g}}{_\mu_\nu}$ to raise and lower indices. However, a simplified index manipulation of $\tensor{\mathcal{Z}}{^\mu_\nu}$ may lead to overlooked contributions. To avoid such omissions, all relevant geometric quantities were derived without any perturbative approximation, leaving the linearization as a final step. The complete expressions are presented in Appendix \ref{App-A}. The components of the linearized Einstein tensor are:

\begin{align}
    a^2 \, \tensor{G}{^\eta_\eta} =& - 3 \, a^2 H^2 \left(1 - 2\, \Phi \right) + 6 \, a H \, \partial_\eta \Psi - 2 \, \nabla^2 \Psi \,, \label{G^eta_eta} \\
    \tensor{G}{^\eta_r} = - \tensor{G}{^r_\eta} =& - \frac{2}{a^2} \, \partial_r \left( a H \Phi + \partial_\eta \Psi \right) , \label{G^eta_r} \\
    \tensor{G}{^\eta_\theta} = -r^2 \, \tensor{G}{^\theta_\eta} =& - \frac{2}{a^2} \, \partial_\theta \left( a H \Phi + \partial_\eta \Psi \right) , \label{G^eta_theta} \\
    \tensor{G}{^\eta_\varphi} = -r^2 \sin^2(\theta) \, \tensor{G}{^\varphi_\eta} =& - \frac{2}{a^2} \, \partial_\varphi \left( a H \Phi + \partial_\eta \Psi \right), \label{G^eta_varphi} \\
    a^2 \, \tensor{G}{^r_r} =& \; \Upsilon \left(1 - 2\, \Phi \right) + 2 \, \partial_\eta^2 \Psi + 2 \, a H \, \partial_\eta \left( \Phi + 2 \, \Psi \right) + {\left(\nabla^2 -  \partial_r^2 \right)} {\left(\Phi - \Psi \right)} \, , \label{G^r_r} \\
    \tensor{G}{^r_\theta} = \, r^2 \, \tensor{G}{^\theta_r} =& \; \frac{r}{a^2} \, \partial_r \left( \frac{\partial_\theta \left(\Psi - \Phi\right)}{r} \right) , \label{G^r_theta} \\
    \tensor{G}{^r_\varphi} = \, r^2 \sin^2(\theta) \, \tensor{G}{^\varphi_r} =& \; \frac{r}{a^2} \, \partial_r \left( \frac{\partial_\varphi \left(\Psi - \Phi\right)}{r} \right) , \label{G^r_varphi} \\
    a^2 \, \tensor{G}{^\theta_\theta} =& \; \Upsilon \left(1 - 2\, \Phi \right) + 2 \, \partial_\eta^2 \Psi + 2 \, a H \, \partial_\eta \left( \Phi + 2 \, \Psi \right) + {\left( \nabla^2 - \frac{\partial_r}{r} - \frac{\partial_\theta^2}{r^2} \right)} {\left(\Phi - \Psi\right)} , \label{G^theta_theta} \\
    \tensor{G}{^\theta_\varphi} = \; \sin^2(\theta) \, \tensor{G}{^\varphi_\theta} =& \; \frac{\sin(\theta)}{a^2 r^2} \, \partial_\theta \left( \frac{\partial_\varphi \left(\Psi - \Phi\right)}{\sin(\theta)} \right) , \label{G^theta_varphi} \\
    a^2 \, \tensor{G}{^\varphi_\varphi} =& \; \Upsilon \left(1 - 2\, \Phi \right) + 2 \, \partial_\eta^2 \Psi + 2 \, a H \, \partial_\eta \left( 
    \Phi + 2 \, \Psi \right) + {\left(\partial_r^2 + \frac{\partial_r}{r}  + \frac{\partial_\theta^2}{r^2} \right)} \left(\Phi - \Psi\right) , \label{G^varphi_varphi}
\end{align}

\noindent
where $\Upsilon(\eta) = a^2 H^2 - \frac{2}{a} \frac{d^2 a}{d \eta^2}$. Besides retaining only terms linear in $\Phi$ and $\Psi$, it is necessary to exclude all contributions dependent on the metric perturbations that are proportional to the spatial derivatives of $\phi$. Upon decomposing $\phi(x) = \overline{\phi}(\eta) + \delta\phi(x)$, such terms are clearly of second order and thus negligible. This consideration is particularly relevant when computing the linearized components of $\tensor{\mathcal{Z}}{^\mu_\nu}$,

\begin{align}
    a^2 \, \tensor{\mathcal{Z}}{^\eta_\eta} =& \; 3 \, \partial_\eta \phi^4 {\left(a H \left(1 - 2\, \Phi \right) - \partial_\eta \Psi \right)} - \nabla^2 \phi^4 \,, \label{tensor_Z^eta_eta} \\
    \tensor{\mathcal{Z}}{^\eta_r} = - \tensor{\mathcal{Z}}{^r_\eta} =& \; \frac{1}{a^2} \left( a H \, \partial_r \phi^4 + \partial_r \Phi \, \partial_\eta \phi^4 - \partial_\eta \partial_r \phi^4 \right) , \label{tensor_Z^eta_r} \\
    \tensor{\mathcal{Z}}{^\eta_\theta} = - r^2 \, \tensor{\mathcal{Z}}{^\theta_\eta} =& \; \frac{1}{a^2} \left( a H \, \partial_\theta \phi^4 + \partial_\theta \Phi \, \partial_\eta \phi^4 - \partial_\eta \partial_\theta \phi^4 \right), \label{tensor_Z^eta_theta} \\
    \tensor{\mathcal{Z}}{^\eta_\varphi} = - r^2 \sin^2(\theta) \, \tensor{\mathcal{Z}}{^\varphi_\eta} =& \; \frac{1}{a^2} \left( a H \, \partial_\varphi \phi^4 + \partial_\varphi \Phi \, \partial_\eta \phi^4 - \partial_\eta \partial_\varphi \phi^4 \right) , \label{tensor_Z^eta_varphi} \\
    a^2 \, \tensor{\mathcal{Z}}{^r_r} =& \; \partial_\eta \phi^4 {\bigg( a H \left(1 - 2\, \Phi \right) - \partial_\eta \Phi - 2 \, \partial_\eta \Psi \bigg)} + \partial_\eta^2 \phi^4 \left(1 - 2\, \Phi \right) - \left( \nabla^2 - \partial_r^2 \right) \phi^4 , \label{tensor_Z^r_r} \\
    \tensor{\mathcal{Z}}{^r_\theta} = r^2 \, \tensor{\mathcal{Z}}{^\theta_r} =& \; \frac{r}{a^2} \, \partial_r \left( \frac{\partial_\theta \phi^4}{r} \right) , \label{tensor_Z^r_theta} \\
    \tensor{\mathcal{Z}}{^r_\varphi} = r^2 \sin^2 (\theta) \, \tensor{\mathcal{Z}}{^\varphi_r} =& \, \frac{r}{a^2} \, \partial_r \left(\frac{\partial_\varphi \phi^4}{r} \right) , \label{tensor_Z^r_varphi} \\
    a^2 \, \tensor{\mathcal{Z}}{^\theta_\theta} =& \; \partial_\eta \phi^4 {\bigg( a H \left(1 - 2\, \Phi \right) - \partial_\eta \Phi - 2 \, \partial_\eta \Psi \bigg)} + \partial_\eta^2 \phi^4 \left(1 - 2\, \Phi \right) - \left( \nabla^2 - \frac{\partial_r}{r} - \frac{\partial_\theta^2}{r^2} \right) \phi^4 , \label{tensor_Z^theta_theta} \\
    \tensor{\mathcal{Z}}{^\theta_\varphi} = \sin^2(\theta) \, \tensor{\mathcal{Z}}{^\varphi_\theta} =& \, \frac{\sin (\theta)}{a^2 r^2}  \, \partial_\theta \left( \frac{\partial_\varphi \phi^4}{\sin (\theta)} \right) , \label{tensor_Z^theta_varphi} \\
    a^2 \, \tensor{\mathcal{Z}}{^\varphi_\varphi} =& \; \partial_\eta \phi^4 {\bigg( a H \left(1 - 2\, \Phi \right) - \partial_\eta \Phi - 2 \, \partial_\eta \Psi \bigg)} + \partial_\eta^2 \phi^4 \left(1 - 2\, \Phi \right) - \left( \partial_r^2 + \frac{\partial_r}{r} + \frac{\partial_\theta^2}{r^2} \right) \phi^4 . \label{tensor_Z^varphi_varphi}
\end{align}

\noindent
Moreover, the stress-energy tensor equals

\begin{equation}
    \tensor{T}{^\mu_\nu} = \tensor{\overline{g}}{^\mu^\alpha} \, \partial_\alpha \phi \partial_\nu \phi + \delta^\mu_\nu \left[\frac{1}{2 a^2} {\left( \partial_\eta \phi \right)}^2 - V(\phi) \right] \,.
    \label{eq::stress_energy^mu_nu}
\end{equation}

\noindent
At last, $\phi(x) = \overline{\phi}(\eta) + \delta\phi(x)$ has to be substituted into $\tensor{T}{^\mu_\nu}$ and $\tensor{\mathcal{Z}}{^\mu_\nu}$, followed by a separation of background and perturbed Einstein equations. This decomposition extends to Eq.~(\ref{eq::KG_eq_curved_non_minimal}). Nonetheless, a transition to the non-relativistic regime is warranted before further proceeding.


\section{THE NON-RELATIVISTIC LIMIT}

In Ref.~\cite{Salehian::2020}, the authors developed a formalism to study the non-relativistic limit within a minimal coupled FLRW metric perturbed solely by scalar modes. Later, this framework was extended in Ref.~\cite{Salehian::2021} to handle arbitrary curved spacetimes and to include self-interactions. In a similar spirit to Refs.~\cite{Namjoo::2017,Cordero::2023}, which focus on the flat spacetime case, the low-energy regime is obtained by mapping the real field $\phi$ to a complex field $\psi$ through an exact transformation. The derivation in the Minkowski spacetime was facilitated by a non-local operator in the field redefinition. However, this method becomes less effective in curved spacetime, as it leads to an equation of motion that differs from the Schrödinger equation \cite{Salehian::2020,Salehian::2021}. Hence, the relation between $\phi$ and $\psi$ now reads:

\begin{align}
    \phi(x) = \frac{1}{\sqrt{2m_a}} \left( e^{-im_at} \, \psi(x) + e^{im_at} \, \psi^*(x) \right) \,, &&
    \partial_t \phi(x) = -i\sqrt{\frac{m_a}{2}} \left( e^{-im_at} \, \psi(x) - e^{im_at} \, \psi^*(x) \right) \,, \label{eq::transf_Salehian}
\end{align}

\noindent
which is no longer a canonical field transformation. Although it may seem natural to express the above relations using conformal time $\eta$ instead of cosmic time $t$, doing so introduces additional terms that complicate the effective description. It follows directly from the above expressions that

\begin{equation}
    e^{-imt} \, \partial_t \psi + e^{imt} \, \partial_t \psi^* = 0 \,. \label{eq::transf_Salehian3}
\end{equation}
    
To simplify the notation, it is convenient to define 

\begin{equation}
    V_\text{int}(\phi) = V(\phi) - \frac{1}{2} m_a^2 \phi^2 + \xi R \, \phi^4 \,,
\end{equation}

\noindent
which collects all self-interaction terms. Moreover, Eq.~(\ref{eq::KG_eq_curved_non_minimal}) can be rewritten as

\begin{equation}
    \frac{1}{\sqrt{-g}} \partial_\mu \left( \sqrt{-g} \tensor{g}{^\mu^\nu} \partial_\nu \phi \right) = \frac{\partial V_\text{int}}{\partial \phi} \,.
    \label{eq::KG_eq_curved_non_minimal2}
\end{equation}

\noindent
Upon expanding the index sum,

\begin{align}
    - \tensor{g}{^\eta^\eta} \partial_\eta^2 \phi =& \, \frac{1}{\sqrt{-g}} \partial_\mu \left( \sqrt{-g} \tensor{g}{^\mu^\eta} \right) \partial_\eta \phi + \frac{1}{\sqrt{-g}} \partial_\mu \left( \sqrt{-g} \tensor{g}{^\mu^i} \right) \partial_i \phi + \tensor{g}{^i^j} \partial_i \partial_j \phi + 2 \tensor{g}{^i^\eta} \partial_i \partial_\eta \phi - m_a^2 \phi - \frac{\partial V_\text{int}}{\partial \phi} \,.
    \label{eq::double_derivative_eta_phi}
\end{align}

\noindent
Meanwhile, from Eq.~(\ref{eq::transf_Salehian}), it is straightforward to verify that

\begin{align}
    \psi(x) = \sqrt{\frac{m_a}{2}} e^{im_at} \left( \phi(x) + \frac{i}{a m_a} \partial_\eta \phi(x) \right) \, .
\end{align}

\noindent
Deriving this equality with respect to $\eta$ leads to

\begin{align}
    i \, \partial_\eta \psi =& - \frac{a m_a}{2} \left( \psi + e^{2im_at} \, \psi^* \right) - \frac{i}{2a} \frac{d a}{d \eta} \left( \psi - e^{2im_at} \, \psi^* \right) - \frac{1}{a \sqrt{2m_a}} e^{im_at} \, \partial_\eta^2 \phi \, .
    \label{eq::motion_psi_non_minimal1}
\end{align}

\noindent
Finally, replacing Eq.~(\ref{eq::double_derivative_eta_phi}) into Eq.~(\ref{eq::motion_psi_non_minimal1}) results in

\begin{align}
    i \tensor{g}{^\eta^\eta} \partial_\eta \psi + \mathcal{D} \, \psi + e^{2im_at} \, \mathcal{D}^* \, \psi^* + \frac{e^{im_at}}{a \sqrt{2m_a}} \frac{\partial V_\text{int}}{\partial \phi} = 0 \, ,
    \label{eq::motion_psi_non_minimal2}
\end{align}

\noindent
where $\mathcal{D}$ is an operator

\begin{align}
    \mathcal{D} \, \psi =& \, \frac{m_a}{2a} \left( a^2 \tensor{g}{^\eta^\eta} + 1 \right) \psi + \frac{i}{2a} \frac{d a}{d \eta} \tensor{g}{^\eta^\eta} \, \psi + \frac{i}{2} \frac{1}{\sqrt{-g}} \partial_\mu \left( \sqrt{-g} \tensor{g}{^\mu^\eta} \right) \psi + i \tensor{g}{^i^\eta} \partial_i \psi - \frac{1}{2a m_a \, \sqrt{-g}} \partial_\mu \left( \sqrt{-g} \tensor{g}{^\mu^i} \right) \partial_i \psi \nonumber \\
    &- \frac{1}{2a m_a} \tensor{g}{^i^j} \, \partial_i \partial_j \psi \, .
\end{align}

\noindent
If all instances of $\eta$ in Eq.~(\ref{eq::motion_psi_non_minimal2}) were changed to $t$, the same equation of motion that was derived in Ref.~\cite{Salehian::2020} would be obtained. However, the definition of $\mathcal{D}$ does differ by a factor of $a$. Nonetheless, the same symbol is retained for the sake of notational simplicity.

Similarly to the flat spacetime case, the rapidly oscillating modes of $\psi$ exert a non-trivial backreaction to the dominant, slowly-varying component. This feedback is represented by the last two terms of Eq.~(\ref{eq::motion_psi_non_minimal2}). As demonstrated in Refs.~\cite{Namjoo::2017,Salehian::2020,Salehian::2021,Cordero::2023}, the fast oscillatory modes induce corrections to the effective potential of $\psi$. Interestingly, the contribution responsible for axion pair-interactions remains unaffected. Therefore, if the axion interactions are reduced to

\begin{align}
    V_{\text{int}} (\phi) = \frac{\lambda_4}{4!} \frac{m_a^2 \, \phi^4}{f_a^2} + \xi R \, \phi^4 \equiv \frac{\Bar{\lambda}_4}{4!} \phi^4 + \xi R \, \phi^4 \, ,
\end{align}

\noindent
the high-frequency contributions can be neglected as a first approximation. For a complete treatment of their impact, see Ref.\cite{Salehian::2020,Salehian::2021}. Consequently, the equation of motion becomes

\begin{align}
    i \tensor{g}{^\eta^\eta} \partial_\eta \psi + \mathcal{D} \, \psi + \frac{1}{8 a m_a^2} \left(\Bar{\lambda}_4 + 24 \xi R \right) \psi |\psi|^2 = 0 \, .
\end{align}

\noindent
Upon inserting the Newtonian gauge metric, multiplying by $\tensor{g}{_\eta_\eta}$, and linearizing,

\begin{align}
    i \partial_\eta \psi =& \; a {\left(m_a \Phi - \frac{3i}{2} H \right)} \psi + \frac{i}{2} \partial_\eta \left( \Phi + 3 \, \Psi \right) \psi - \frac{\nabla^2 \psi}{2am_a} + \left( \frac{a \Bar{\lambda}_4}{8 m_a^2} + \frac{9\xi}{a m_a^2} (a^2 H^2 - \Upsilon) \right) \psi |\psi|^2 \nonumber \\ 
    &- \frac{6\xi}{a m_a^2} \left( 3 a H \, \partial_\eta \left( \Phi + 3 \, \Psi \right) + 3 \, \partial_\eta^2 \Psi + \nabla^2 \Phi - 2 \, \nabla^2 \Psi \right) \psi |\psi|^2 \,,
    \label{eq::motion_psi_non_minimal3}
\end{align}

\noindent
where the $\Bar{\lambda}_4 \Phi$ term was neglected as $\Bar{\lambda}_4 = \lambda_4 m_a^2/f_a^2 \ll 1$ for $m_a^2/f_a^2 \in (10^{-52} - 10^{-44})$. The background dynamics are obtained by substituting $\psi(x) = \overline{\psi} (\eta)$ and setting the metric perturbations to zero,

\begin{align}
    i \partial_\eta \overline{\psi} =& - \frac{3i}{2} a H \, \overline{\psi} + \left( \frac{a \Bar{\lambda}_4}{8 m_a^2}  + \frac{9\xi}{a m_a^2} (a^2 H^2 - \Upsilon) \right) \overline{\psi} |\overline{\psi}|^2\,.
    \label{eq::motion_psi_background}
\end{align}

\noindent
Next, the non-relativistic field is split as $\psi(x) = \overline{\psi} (\eta) + \delta \psi(x)$ with $|\delta \psi| \ll |\overline{\psi}|$. To eliminate ambiguity in the decomposition, it is common to impose that the spatial average of $\delta \psi$ vanishes \cite{Salehian::2020}. Inserting this ansatz into Eq.~(\ref{eq::motion_psi_non_minimal3}) and subtracting the background evolution yields

\begin{align}
    i \partial_\eta (\delta\psi) =& \; a m_a \Phi \, \overline{\psi} - \frac{3i}{2} a H \, \delta \psi  + \frac{i}{2} \partial_\eta \left( \Phi + 3 \, \Psi \right) \overline{\psi} - \frac{\nabla^2 (\delta \psi)}{2am_a} + \left( \frac{a \Bar{\lambda}_4}{8 m_a^2}  + \frac{9\xi}{a m_a^2} (a^2 H^2 - \Upsilon) \right) \left( \overline{\psi}^2 \delta \psi^* + 2 \, |\overline{\psi}|^2 \delta \psi \right) \nonumber \\ 
    &- \frac{6\xi}{a m_a^2} \left( 3 \, a H \, \partial_\eta \left(\Phi + 3 \, \Psi \right) + 3 \, \partial_\eta^2 \Psi + \nabla^2 \Phi - 2 \, \nabla^2 \Psi \right) \overline{\psi} |\overline{\psi}|^2 \,.
    \label{eq::motion_psi_perturbation}
\end{align}

Afterwards, the non-relativistic limit of the modified Einstein equations has to be obtained. Inserting Eq.~(\ref{eq::transf_Salehian}) into Eq.~(\ref{eq::stress_energy^mu_nu}) leads to the following non-zero components of the stress-energy tensor:

\begin{align}
    \tensor{T}{^\eta_\eta} =& - m_a |\psi|^2 - \frac{\Bar{\lambda}_4}{16 m_a^2} |\psi|^4 \,, \\
    \tensor{T}{^\eta_r} = - \tensor{T}{^r_\eta} =& \; \frac{i}{2a} {\left(\psi \, \partial_r \psi^* - \psi^* \partial_r \psi \right)} \,, \\
    \tensor{T}{^\eta_\theta} = - r^2 \, \tensor{T}{^\theta_\eta} =& \; \frac{i}{2a} {\left(\psi \, \partial_\theta \psi^* - \psi^* \partial_\theta \psi \right)} \,, \\
    \tensor{T}{^\eta_\varphi} = - r^2 \sin^2 (\theta) \, \tensor{T}{^\varphi_\eta} =& \; \frac{i}{2a} {\left(\psi \, \partial_\varphi \psi^* - \psi^* \partial_\varphi \psi \right)} \,, \\
    \tensor{T}{^r_r} = \tensor{T}{^\theta_\theta} = \tensor{T}{^\varphi_\varphi} =& - \frac{\Bar{\lambda}_4}{16 m_a^2} |\psi|^4 \,.
\end{align}

\noindent
It is also necessary to express $\tensor{\mathcal{Z}}{^\mu_\nu}$ in terms of $\psi$. Using Eqs.~(\ref{eq::transf_Salehian})~-~(\ref{eq::transf_Salehian3}), it is possible to show that

\begin{align}
    \partial_\eta \phi^4 \thickapprox 0\,, && \partial_\eta \partial_i \phi^4 \thickapprox 0\,, && \partial_\eta^2 \phi^4 \thickapprox \frac{6ia}{m_a} |\psi|^2 \left(\psi \, \partial_\eta \psi^* - \psi^* \partial_\eta \psi \right)\,.
\end{align}

\noindent
Additionally, any term involving a double spatial derivative of $\phi^4$ is evaluated by replacing $\phi^4 \thickapprox \frac{3}{2m_a^2} |\psi|^4$. It is advantageous to use Eq.~(\ref{eq::motion_psi_non_minimal3}) to eliminate the $\partial_\eta \psi$ dependence inside $\partial_\eta^2 \phi^4$. This leads to

\begin{align}
    \frac{\partial_\eta^2 \phi^4}{a^2} \left(1 - 2\, \Phi \right) \thickapprox& \quad \frac{3 |\psi|^2}{a^2 m_a^2} \left(\psi \nabla^2 \psi^* + \psi^* \nabla^2 \psi \right) - 12 \, \Phi \, |\psi|^4 - \left(\frac{3 \Bar{\lambda}_4}{2m_a^3} + \frac{108 \xi}{a^2 m_a^3} \left(a^2 H^2 - \Upsilon \right) \left(1 - 2\, \Phi \right) \right) |\psi|^6 \nonumber \\
    &+ \frac{72 \xi}{a^2 m_a^3} \left( 3 a H \, \partial_\eta \left( \Phi + 3 \, \Psi \right) + 3 \, \partial_\eta^2 \Psi + \nabla^2 \Phi - 2 \, \nabla^2 \Psi \right) |\psi|^6 \,.
\end{align}

\noindent
When $\tensor{\mathcal{Z}}{^\mu_\nu}$ is inserted into Eq.~(\ref{eq::Einstein_field_eqs_non_minimal}), the last three terms of the above expression act as corrections to the axion potential. Given the current simplification to pairwise interactions, the $|\psi|^6$ terms are disregarded because their influence is incomplete without including the fast-oscillating corrections and the $\phi^6$ term of $V(\phi)$. 

The background modified Einstein equations are obtained by replacing $\psi(x) = \overline{\psi} (\eta)$ and $\Phi(x) = \Psi(x) = 0$ in Eq.~(\ref{eq::Einstein_field_eqs_non_minimal}). The only non-trivial equations are the $\eta \eta$ component,

\begin{equation}
    \frac{3 H^2}{8\pi G} - m_a |\overline{\psi}|^2 - \left( \frac{\Bar{\lambda}_4}{16 m_a^2} + \frac{9 \xi H^2}{m_a^2} \right) |\overline{\psi}|^4 = 0 \,, \label{eq::Einstein_field_eqs_background^eta_eta}
\end{equation}

\noindent
and the diagonal components of the spatial sector,

\begin{equation}
    \frac{1}{8\pi G} \frac{\Upsilon}{a^2} + \left( \frac{\Bar{\lambda}_4}{16 m_a^2} - \frac{3 \xi \Upsilon}{a^2 m_a^2} \right) |\overline{\psi}|^4 = 0 \,.
    \label{eq::Einstein_field_eqs_background^r_r}
\end{equation}

\noindent
While Eqs.~(\ref{eq::Einstein_field_eqs_background^eta_eta}) and~(\ref{eq::Einstein_field_eqs_background^r_r}) govern the evolution of $a$, they reflect a simplified scenario in which only $\overline{\psi}$ contributes to the energy content. A realistic cosmological model must include radiation, baryonic matter, dark matter, and dark energy. In practice, $a$ is treated as a known function, obtained from observationally supported cosmological models.

The perturbed form of Eq.~(\ref{eq::Einstein_field_eqs_non_minimal}) is obtained by inserting $\psi(x) = \overline{\psi} (\eta) + \delta \psi(x)$, followed by the removal of their background counterparts. Notably, $\Phi(x)$ and $\Psi(x)$ are not dynamical variables but instead serve as auxiliary fields. Hence, the perturbed Einstein equations act as constraints rather than equations of motion. These ligadures are best written in terms of

\begin{align}
    \mathcal{R}_{\delta \psi} =& \; \overline{\psi} \, \delta \psi^* + \overline{\psi}^* \delta \psi \,, && \mathcal{I}_{\delta \psi} = i \left( \overline{\psi} \, \delta \psi^* - \overline{\psi}^* \delta \psi \right), \nonumber \\
    8\pi \, \mathcal{G}(\eta) =& \; {\left( \frac{1}{8\pi G} - \frac{3 \xi |\overline{\psi}|^4}{m_a^2} \right)}^{-1} .
\end{align} 

\noindent
The $\eta \eta$ component is

\begin{align}
    \frac{1}{4\pi \mathcal{G}} \left(3 \, a^2 H^2 \Phi + 3 \, aH \, \partial_\eta \Psi - \nabla^2 \Psi \right) + a^2 \left[ m_a + \frac{|\overline{\psi}|^2}{m_a^2} \left( \frac{\Bar{\lambda}_4}{8} + \frac{6 \xi}{a^2} \left(3 \, a^2 H^2 - \nabla^2 \right) \right) \right] \mathcal{R}_{\delta \psi} = 0 \,. \label{eq::Einstein_field_eqs_pert^eta_eta}
\end{align}

\noindent
The temporal-spatial components can be combined into

\begin{align}
    \frac{1}{4\pi \mathcal{G}} \nabla \left( a H \Phi + \partial_\eta \Psi \right) =& \; \frac{6 \xi a H |\overline{\psi}|^2}{m_a^2} \nabla \mathcal{R}_{\delta \psi} - \frac{a}{2} \nabla \mathcal{I}_{\delta \psi}  \,. \label{eq::Einstein_field_eqs_pert^eta_r}
\end{align}

\noindent
The trace of the spatial sector equals

\begin{align}
    \frac{1}{8\pi \mathcal{G}} \left( \partial_\eta^2 \Psi + \frac{1}{3} \, \nabla^2 (\Phi - \Psi) + a H \, \partial_\eta \left( \Phi + 2 \, \Psi \right) - \Upsilon \, \Phi\right) -12 \xi a^2 |\overline{\psi}|^4 \, \Phi + \frac{a^2 |\overline{\psi}|^2}{m_a^2} \left( \frac{\Bar{\lambda}_4}{16} - \frac{\xi}{a^2} \left(3 \Upsilon - \nabla^2 \right) \right) \mathcal{R}_{\delta \psi} = 0 \,. \label{eq::Einstein_field_eqs_pert-spatial_trace}
\end{align}

\noindent
The non-diagonal space Einstein equations are

\begin{align}
    \frac{1}{8\pi \mathcal{G}} \, \partial_r \left( \frac{\partial_\theta {\left(\Phi - \Psi\right)}}{r} \right) =& \; \frac{6 \xi |\overline{\psi}|^2}{m_a^2} \, \partial_r \left( \frac{\partial_\theta \mathcal{R}_{\delta \psi}}{r}\right) \,, \label{eq::Einstein_field_eqs_pert^r_eta} \\
    \frac{1}{8\pi \mathcal{G}} \, \partial_r \left( \frac{\partial_\varphi {\left(\Phi - \Psi\right)}}{r} \right) =& \; \frac{6 \xi |\overline{\psi}|^2}{m_a^2} \, \partial_r \left( \frac{\partial_\varphi \mathcal{R}_{\delta \psi}}{r} \right) \,, \label{eq::Einstein_field_eqs_pert^r_theta} \\
    \frac{1}{8\pi \mathcal{G}} \, \partial_\theta \left( \frac{\partial_\varphi {\left(\Phi - \Psi\right)}}{\sin(\theta)} \right) =& \: \frac{6 \xi |\overline{\psi}|^2}{m_a^2}  \, \partial_\theta \left( \frac{\partial_\varphi \mathcal{R}_{\delta \psi}}{\sin (\theta)} \right) \,. \label{eq::Einstein_field_eqs_pert^r_varphi}
\end{align}

\noindent
Imposing $a = 1$, $\xi = 0$, $\Phi = \Psi$, and $\partial_\eta \Phi = 0$ in Eqs.~(\ref{eq::motion_psi_non_minimal3}) and~(\ref{eq::Einstein_field_eqs_pert^eta_eta}) reduces the system to the Gross-Pitaevskii-Poisson equations. Further omitting the axion self-interactions yields the Schrödinger–Poisson equations. Both prescriptions are often used to model the non-relativistic dynamics of gravitationally interacting axions. Thus, the equations derived in this work can be viewed as a systematic refinement of these models.

It is noteworthy that if $\delta \psi(x)$ remains spherically symmetric throughout its entire evolution, Eqs.~(\ref{eq::Einstein_field_eqs_pert^r_eta})~-~(\ref{eq::Einstein_field_eqs_pert^r_varphi}) naturally enforce $\Psi = \Phi$, which mirrors an absence of anisotropic stress. Under this simplification, Eq.~(\ref{eq::motion_psi_perturbation}) reduces to

\begin{align}
    i \partial_\eta (\delta\psi) =& \; a m_a \Phi \, \overline{\psi} - \frac{3i}{2} a H \, \delta \psi  + 2i \, \partial_\eta \Phi \, \overline{\psi} - \frac{\nabla^2 (\delta \psi)}{2am_a} + \left( \frac{a \Bar{\lambda}_4}{8 m_a^2}  + \frac{9\xi}{a m_a^2} (a^2 H^2 - \Upsilon) \right) \left( \overline{\psi}^2 \delta \psi^* + 2 \, |\overline{\psi}|^2 \delta \psi \right) \nonumber \\ 
    &- \frac{6\xi}{a m_a^2} \left( 12 \, a H \, \partial_\eta \Phi + 3 \, \partial_\eta^2 \Phi - \, \nabla^2 \Phi \right) \overline{\psi} |\overline{\psi}|^2 \,.
    \label{eq::motion_psi_perturbation_simplified}
\end{align}

\noindent
Moreover, Eqs.~(\ref{eq::Einstein_field_eqs_pert^eta_eta})~-~(\ref{eq::Einstein_field_eqs_pert-spatial_trace}) become

\begin{gather}
    \frac{1}{4\pi \mathcal{G}} \left(3 \, a^2 H^2 \Phi + 3 \, aH \, \partial_\eta \Phi - \nabla^2 \Phi \right) + a^2 \left[ m_a + \frac{|\overline{\psi}|^2}{m_a^2} \left( \frac{\Bar{\lambda}_4}{8} + \frac{6 \xi}{a^2} \left(3 \, a^2 H^2 - \nabla^2 \right) \right) \right] \mathcal{R}_{\delta \psi} = 0 \,. \label{eq::Einstein_field_eqs_pert^eta_eta_simplified} \\
    \frac{1}{4\pi \mathcal{G}} \nabla \left( a H \Phi + \partial_\eta \Phi \right) = \; \frac{6 \xi a H |\overline{\psi}|^2}{m_a^2} \nabla \mathcal{R}_{\delta \psi} - \frac{a}{2} \nabla \mathcal{I}_{\delta \psi}  \,. \label{eq::Einstein_field_eqs_pert^eta_r_simplified} \\
    \frac{1}{8\pi \mathcal{G}} \left( \partial_\eta^2 \Phi + 3 \, a H \, \partial_\eta \Phi - \Upsilon \, \Phi\right) -12 \xi a^2 |\overline{\psi}|^4 \, \Phi + \frac{a^2 |\overline{\psi}|^2}{m_a^2} \left( \frac{\Bar{\lambda}_4}{16} - \frac{\xi}{a^2} \left(3 \Upsilon - \nabla^2 \right) \right) \mathcal{R}_{\delta \psi} = 0 \,. \label{eq::Einstein_field_eqs_pert-spatial_trace_simplified}
\end{gather}

Finally, the covariant continuity equations are often relevant in the dynamics of gravitational systems. The condition $\nabla_ \mu \tensor{T}{^\mu_\nu} = 0$ is a direct consequence of Einstein field equations due to contracted Bianchi identity $\nabla_ \mu \tensor{G}{^\mu^\nu} = 0$. However, the non-minimal coupling present in this work modifies the Einstein equations, yielding

\begin{equation}
    \nabla_\mu \tensor{T}{^\mu_\nu} + 2 \xi \, \partial_\mu \phi^4 \, \tensor{G}{^\mu_\nu} = 0 \,.
\end{equation}

\noindent
In principle, this relation can be expanded and expressed in terms of $\psi$ to obtain the non-relativistic limit. However, these equations are automatically satisfied by any solution to Eq.~(\ref{eq::Einstein_field_eqs_non_minimal}) and serve as consistency conditions rather than new constraints. For this reason, they were not explicitly evaluated in this work.


\section{FROM GROSS-PITAEVSKII TO SCHRÖDINGER}

Even though $\psi(x) = \overline{\psi}(\eta) + \delta \psi(x)$ is a convenient way to separate the background from the perturbation, the full field is still $\psi(x)$ and Eq.~(\ref{eq::motion_psi_non_minimal3}) continues to describe its dynamics. This equation can be recast as

\begin{equation}
    i \partial_\eta \psi = - \frac{1}{2ma} \nabla^2 \psi + \mathcal{U}\, \psi + \mathcal{U}_0 \, \psi |\psi|^2 \,, \label{eq::Gross-Pitaevskii_modified}
\end{equation}

\noindent
where

\begin{align}
    \mathcal{U}(x) =& \; a m_a \Phi - \frac{3ia}{2}H + \frac{i}{2} \partial_\eta \left(\Phi + 3 \, \Psi \right) \,, \qquad
    \mathcal{U}_0(x) = \; \frac{a \Bar{\lambda}_4}{8 m_a^2} + \frac{3\xi g(x)}{a m_a^2} \,, \nonumber \\
    g(x) =& \; 3\,a^2 H^2 - 3 \Upsilon -6 \, a H \, \partial_\eta \left( \Phi + 3 \, \Psi \right) - 6 \, \partial_\eta^2 \Psi - 2 \, \nabla^2 \Phi + 4 \, \nabla^2 \Psi \,.
\end{align}

\noindent
This expression closely resembles the time-dependent Gross–Pitaevskii equation,

\begin{equation}
    i \partial_t \psi = - \frac{1}{2m} \nabla^2 \psi + \text{U}\, \psi + \text{U}_0 \, \psi |\psi|^2 \,, \label{eq::Gross-Pitaevskii}
\end{equation}

\noindent
where U$(x)$ is an external potential and U$_0$ is a constant that characterizes the $s$-wave scattering between two bosons. Eq.~(\ref{eq::Gross-Pitaevskii}) emerges from the Hartree-Fock treatment of a system of $N$ bosons near the ground state. Hence, it is insightful to apply the inverse procedure to Eq.~(\ref{eq::Gross-Pitaevskii_modified}) to recover the $N$-body Schrödinger equation.

Considering only variations with respect to $\psi$ and $\psi^*$, the action that corresponds to Eq.~(\ref{eq::Gross-Pitaevskii_modified}) is

\begin{align}
    S = \int d\eta \int_V d\Vec{r} \;  \bigg[& i \psi^*{(\eta,\Vec{r})} \, \partial_\eta \psi(\eta,\Vec{r}) + \frac{1}{2ma} \psi^*(\eta,\Vec{r}) \, \nabla^2 \psi(\eta,\Vec{r}) - \mathcal{U}(\eta,\Vec{r}) \, |\psi(\eta,\Vec{r})|^2 - \frac{1}{2} \mathcal{U}_0(\eta,\Vec{r}) \, |\psi(\eta,\Vec{r})|^4 \bigg].
\end{align}

\noindent
To avoid infinities, the spatial integration is done over a finite volume $V$. The system is subject to

\begin{equation}
    \frac{1}{V} \int_V d\Vec{r} \; |\psi(\eta,\Vec{r})|^2 = |\overline{\psi}(\eta)|^2 \,,
\end{equation}

\noindent
which can be computed explicitly by solving Eq.~(\ref{eq::motion_psi_background}). The wavefunction $\chi$ of the whole system is modeled as 

\begin{align}
     \chi(\eta,\Vec{r}_1, \,\ldots\,, \Vec{r}_N) = \frac{1}{|\overline{\psi}|^{N-1}} \prod_{k=1}^N \psi(\eta,\Vec{r}_{k}) \,,
\end{align}

\noindent
since it ensures that 

\begin{equation}
    \left(\prod_{j=1}^N \frac{1}{V} \int_V d\Vec{r}_j \right) |\chi|^2 = |\overline{\psi}(\eta)|^2 \,.
\end{equation}

\noindent
A straightforward calculation shows that

\begin{align}
    S = \frac{1}{N V^{N-1}} \int d\eta  \left(\prod_{j=1}^N \int_V d\Vec{r}_j \right) \chi^* \left( i \partial_\eta + \sum_{k=1}^N \frac{\nabla_k^2}{2ma} - \text{U}_\text{ext} - \text{U}_\text{int} \right) \chi\,,
    \label{eq::action_chi}
\end{align}

\noindent
where

\begin{align}
    \text{U}_\text{ext} = \; \sum_{k=1}^N \left( \mathcal{U}(\eta,\Vec{r}_k) - \frac{i(N-1)}{N^2} \partial_\eta \ln |\overline{\psi}| \right) \,, \qquad
    \text{U}_{\text{int}} = \; \frac{V |\overline{\psi}|^2}{2(N-1)} \; \sum_{k=1}^N \sum_{l \neq k}^N \; \mathcal{U}_0(\eta,\Vec{r}_k) \, \delta(\Vec{r}_k - \Vec{r}_l) \,.
\end{align}

\noindent
Eq.~(\ref{eq::action_chi}) directly leads to

\begin{align}
    i \partial_\eta \chi = \left( - \sum_{k=1}^N \frac{\nabla_k^2}{2ma} + \text{U}_\text{ext} + \text{U}_\text{int}\right) \, \chi \,.
\end{align}

\noindent
Therefore, the Hamiltonian operator that encapsulates the system’s dynamics is

\begin{align}
    \Hat{H} =&\; - \sum_{k=1}^N \frac{\nabla_k^2}{2ma} + \sum_{k=1}^N \left( \mathcal{U}(\eta,\Vec{r}_k) - \frac{i(N-1)}{N^2} \partial_\eta \ln |\overline{\psi}| \right) + \frac{V |\overline{\psi}|^2}{2(N-1)} \; \sum_{k=1}^N \sum_{l \neq k}^N \; \mathcal{U}_0(\eta,\Vec{r}_k) \; \delta(\Vec{r}_k - \Vec{r}_l) \,.
\end{align}

\noindent
This approach is particularly useful, as it enables a future computation of the partition function for a system of axions subject to gravitationally mediated pairwise interactions.

\section{CONCLUSIONS}
The novel formulation discussed in this work may yield new insights on the gravitational collapse underlying axion miniclusters and axion stars. However, a key limitation with the proposed framework lies in the unknown value of $\xi$. This parameter can be inferred by computing the axion–axion scattering cross section with non-minimal coupling and comparing the result to the DM-DM cross section discussed in Ref.~\cite{ParticleDataGroup::2022}. Since gravitational effects only play an important role at high axion densities, direct laboratory measurements are unfeasible. Consequently, astrophysical environments, such as galaxy cluster mergers, must be used. However, modeling the axion distribution within these systems is non-trivial, especially in the presence of substructures. To ensure that the axion’s cosmological history remains unaffected,  the order of magnitude of $\xi R\phi^4$ should not exceed that of $\Bar{\lambda}_4 \phi^4/4! $. This places a natural constraint on the value of $\xi$.

While the system of equations developed in this work can, in principle, be solved numerically to obtain a detailed description of the dynamics, a number of important properties can be understood analytically. This can be accomplished by reformulating the evolution equations as hydrodynamic equations through the Madelung transformation, mirroring the investigations done in Ref.\cite{Chavanis::2011,Chavanis::2016}. The results of these publications can subsequently be revisited with the inclusion of the non-minimal coupling term to assess its impact on the system.


\begin{acknowledgments}
We acknowledge financial support from 
Ministerio Espa\~nol de Ciencia e Innovaci\'on under grant No. PID2022-140440NB-C22;
Junta de Andaluc\'ia under contract No. FQM-370 as well as PCI+D+i under the title: ``Tecnolog\'\i as avanzadas para la exploraci\'on del universo y sus componentes" (Code AST22-0001).
\end{acknowledgments}


\bibliography{print_AxionGravityCoupling}

@article{Battat::2024,
    author = {Battat, James B. R.},
    title = {Resource Letter DM1: Dark matter: An overview of theory and experiment},
    journal = {American Journal of Physics},
    volume = {92},
    number = {4},
    pages = {247-257},
    year = {2024},
    month = {04},
    issn = {0002-9505},
    doi = {10.1119/5.0187346},
    url = {https://doi.org/10.1119/5.0187346},
    eprint = {https://pubs.aip.org/aapt/ajp/article-pdf/92/4/247/20162494/247\_1\_5.0187346.pdf},
}

@article{Peccei::1977,
  title = {$\mathrm{CP}$ Conservation in the Presence of Pseudoparticles},
  author = {Peccei, R. D. and Quinn, Helen R.},
  journal = {Phys. Rev. Lett.},
  volume = {38},
  issue = {25},
  pages = {1440--1443},
  numpages = {0},
  year = {1977},
  month = {Jun},
  publisher = {American Physical Society},
  doi = {10.1103/PhysRevLett.38.1440},
  url = {https://link.aps.org/doi/10.1103/PhysRevLett.38.1440}
}

@book{Jackson::2023,
    author = "Derek F. Jackson Kimball and Karl van Bribber",
    title = "The Search for Ultralight Bosonic Dark Matter",
    publisher = "Springer" ,
    year = "2023"
}

@article{ParticleDataGroup::2022,
    author = "Workman, R. L. and others",
    collaboration = "Particle Data Group",
    title = "{Review of Particle Physics}",
    doi = "10.1093/ptep/ptac097",
    journal = "PTEP",
    volume = "2022",
    pages = "083C01",
    year = "2022"
}

@misc{Irastorza::2021,
      title={An introduction to axions and their detection}, 
      author={Igor G. Irastorza},
      howpublished = {SciPost Physics Lecture Notes, Les Houches Summer School Series},
      year={2021},
      eprint={2109.07376},
      archivePrefix={arXiv},
      primaryClass={hep-ph}
}

@article{Luzio::2020,
   title={The landscape of QCD axion models},
   volume={870},
   ISSN={0370-1573},
   url={http://dx.doi.org/10.1016/j.physrep.2020.06.002},
   DOI={10.1016/j.physrep.2020.06.002},
   journal={Physics Reports},
   publisher={Elsevier BV},
   author={Di Luzio, Luca and Giannotti, Maurizio and Nardi, Enrico and Visinelli, Luca},
   year={2020},
   month={Jul},
   pages={1–117} 
}

@article{Peccei_2::1977,
  title = {Constraints imposed by $\mathrm{CP}$ conservation in the presence of pseudoparticles},
  author = {Peccei, R. D. and Quinn, Helen R.},
  journal = {Phys. Rev. D},
  volume = {16},
  issue = {6},
  pages = {1791--1797},
  numpages = {0},
  year = {1977},
  month = {Sep},
  publisher = {American Physical Society},
  doi = {10.1103/PhysRevD.16.1791},
  url = {https://link.aps.org/doi/10.1103/PhysRevD.16.1791}
}

@inproceedings{snowmass_axions::2022,
    author = "Adams, C. B. and others",
    title = "{Axion Dark Matter}",
    booktitle = "{Snowmass 2021}",
    eprint = "2203.14923",
    archivePrefix = "arXiv",
    primaryClass = "hep-ex",
    reportNumber = "FERMILAB-CONF-22-996-PPD-T",
    month = "3",
    year = "2022"
}

@article{Braaten::2019,
  title = {Colloquium: The physics of axion stars},
  author = {Braaten, Eric and Zhang, Hong},
  journal = {Rev. Mod. Phys.},
  volume = {91},
  issue = {4},
  pages = {041002},
  numpages = {22},
  year = {2019},
  month = {Oct},
  publisher = {American Physical Society},
  doi = {10.1103/RevModPhys.91.041002},
  url = {https://link.aps.org/doi/10.1103/RevModPhys.91.041002}
}

@article{Namjoo::2017,
    author = "Namjoo, Mohammad Hossein and Guth, Alan H. and Kaiser, David I.",
    title = "{Relativistic Corrections to Nonrelativistic Effective Field Theories}",
    eprint = "1712.00445",
    archivePrefix = "arXiv",
    primaryClass = "hep-ph",
    reportNumber = "MIT-CTP-4955, MIT-CTP/4955",
    doi = "10.1103/PhysRevD.98.016011",
    journal = "Phys. Rev. D",
    volume = "98",
    number = "1",
    pages = "016011",
    year = "2018"
}

@article{Cordero::2023,
    author = "Cordero-Patino, Bryan and Duenas-Vidal, \'Alvaro and Segovia, Jorge",
    title = "{Higher-Order Corrections to the Effective Field Theory of Low-Energy Axions}",
    eprint = "2306.08721",
    archivePrefix = "arXiv",
    primaryClass = "hep-ph",
    doi = "10.3390/sym15122098",
    journal = "Symmetry",
    volume = "15",
    number = "12",
    pages = "2098",
    year = "2023"
}

@book{Padmanabhan::2010,
    author = {{Padmanabhan}, T.},
    title = "{Gravitation: Foundations and Frontiers}",
    year = 2010,
    adsurl = {https://ui.adsabs.harvard.edu/abs/2010grav.book.....P},
    publisher = "{Cambridge University Press}",
}

@book{Leonard::2009, 
    place={Cambridge}, 
    edition={1}, 
    title={Quantum Field Theory in Curved Spacetime},
    publisher={Cambridge University Press}, 
    author={P. Leonard and D. Toms},
    year={2009}
}

@book{Gorbunov::2011,
    author = "Gorbunov, Dmitry S. and Rubakov, Valery A.",
    title = "{Introduction to the theory of the early universe: Cosmological perturbations and inflationary theory}",
    doi = "10.1142/7873",
    year = "2011",
    publisher = "World Scientific"
}

@article{Sotiriou::2010,
   title={f(R) theories of gravity},
   volume={82},
   ISSN={1539-0756},
   url={http://dx.doi.org/10.1103/RevModPhys.82.451},
   DOI={10.1103/revmodphys.82.451},
   number={1},
   journal={Reviews of Modern Physics},
   publisher={American Physical Society (APS)},
   author={Sotiriou, Thomas P. and Faraoni, Valerio},
   year={2010},
   month=mar, pages={451–497} }

@article{Bezrukov::2008,
   title={The Standard Model Higgs boson as the inflaton},
   volume={659},
   ISSN={0370-2693},
   url={http://dx.doi.org/10.1016/j.physletb.2007.11.072},
   DOI={10.1016/j.physletb.2007.11.072},
   number={3},
   journal={Physics Letters B},
   publisher={Elsevier BV},
   author={Bezrukov, Fedor and Shaposhnikov, Mikhail},
   year={2008},
   month=jan, pages={703–706} }

@article{Hwang::2000,
   title={Conserved cosmological structures in the one-loop superstring effective action},
   volume={61},
   ISSN={1089-4918},
   url={http://dx.doi.org/10.1103/PhysRevD.61.043511},
   DOI={10.1103/physrevd.61.043511},
   number={4},
   journal={Physical Review D},
   publisher={American Physical Society (APS)},
   author={Hwang, Jai-chan and Noh, Hyerim},
   year={2000},
   month=jan }

@article{Salehian::2020,
    author = "Salehian, Borna and Namjoo, Mohammad Hossein and Kaiser, David I.",
    title = "{Effective theories for a nonrelativistic field in an expanding universe: Induced self-interaction, pressure, sound speed, and viscosity}",
    eprint = "2005.05388",
    archivePrefix = "arXiv",
    primaryClass = "astro-ph.CO",
    reportNumber = "Preprint MIT-CTP/5203",
    doi = "10.1007/JHEP07(2020)059",
    journal = "JHEP",
    volume = "07",
    pages = "059",
    year = "2020"
}

@article{Salehian::2021,
    author = "Salehian, Borna and Zhang, Hong-Yi and Amin, Mustafa A. and Kaiser, David I. and Namjoo, Mohammad Hossein",
    title = {{Beyond Schr\"odinger-Poisson: nonrelativistic effective field theory for scalar dark matter}},
    eprint = "2104.10128",
    archivePrefix = "arXiv",
    primaryClass = "astro-ph.CO",
    reportNumber = "MIT-CTP/5293",
    doi = "10.1007/JHEP09(2021)050",
    journal = "JHEP",
    volume = "09",
    pages = "050",
    year = "2021"
}

@misc{Chang::2024,
      title={Axion Stars: Mass Functions and Constraints}, 
      author={Jae Hyeok Chang and Patrick J. Fox and Huangyu Xiao},
      year={2024},
      eprint={2406.09499},
      archivePrefix={arXiv},
      primaryClass={hep-ph},
      url={https://arxiv.org/abs/2406.09499}, 
}

@article{Zhang::2018,
    author = "Zhang, Hong",
    title = "{Axion Stars}",
    eprint = "1810.11473",
    archivePrefix = "arXiv",
    primaryClass = "hep-ph",
    doi = "10.3390/sym12010025",
    journal = "Symmetry",
    volume = "12",
    number = "1",
    pages = "25",
    year = "2019"
}

@article{Jackson_Kimball::2018,
   title={Searching for axion stars and Q-balls with a terrestrial magnetometer network},
   volume={97},
   ISSN={2470-0029},
   url={http://dx.doi.org/10.1103/PhysRevD.97.043002},
   DOI={10.1103/physrevd.97.043002},
   number={4},
   journal={Physical Review D},
   publisher={American Physical Society (APS)},
   author={Jackson Kimball, D.F. and Budker, D. and Eby, J. and Pospelov, M. and Pustelny, S. and Scholtes, T. and Stadnik, Y.V. and Weis, A. and Wickenbrock, A.},
   year={2018},
   month=feb }

@article{Eby::2016,
   title={Boson stars from self-interacting dark matter},
   volume={2016},
   ISSN={1029-8479},
   url={http://dx.doi.org/10.1007/JHEP02(2016)028},
   DOI={10.1007/jhep02(2016)028},
   number={2},
   journal={Journal of High Energy Physics},
   publisher={Springer Science and Business Media LLC},
   author={Eby, Joshua and Kouvaris, Chris and Nielsen, Niklas Grønlund and Wijewardhana, L. C. R.},
   year={2016},
   month=feb }

@misc{Ilie::2025,
      title={Spectroscopic Supermassive Dark Star candidates}, 
      author={Cosmin Ilie and Sayed Shafaat Mahmud and Jillian Paulin and Katherine Freese},
      year={2025},
      eprint={2505.06101},
      archivePrefix={arXiv},
      primaryClass={astro-ph.CO},
      url={https://arxiv.org/abs/2505.06101}, 
}

@article{Bai::2016,
   title={Hydrogen axion star: metallic hydrogen bound to a QCD axion BEC},
   volume={2016},
   ISSN={1029-8479},
   url={http://dx.doi.org/10.1007/JHEP12(2016)127},
   DOI={10.1007/jhep12(2016)127},
   number={12},
   journal={Journal of High Energy Physics},
   publisher={Springer Science and Business Media LLC},
   author={Bai, Yang and Barger, Vernon and Berger, Joshua},
   year={2016},
   month=dec }

@article{Wu::2022,
   title={Dark stars powered by self-interacting dark matter},
   volume={106},
   ISSN={2470-0029},
   url={http://dx.doi.org/10.1103/PhysRevD.106.043028},
   DOI={10.1103/physrevd.106.043028},
   number={4},
   journal={Physical Review D},
   publisher={American Physical Society (APS)},
   author={Wu, Youjia and Baum, Sebastian and Freese, Katherine and Visinelli, Luca and Yu, Hai-Bo},
   year={2022},
   month=aug }

@article{Chavanis::2011,
   title={Mass-radius relation of Newtonian self-gravitating Bose-Einstein condensates with short-range interactions. I. Analytical results},
   volume={84},
   ISSN={1550-2368},
   url={http://dx.doi.org/10.1103/PhysRevD.84.043531},
   DOI={10.1103/physrevd.84.043531},
   number={4},
   journal={Physical Review D},
   publisher={American Physical Society (APS)},
   author={Chavanis, Pierre-Henri},
   year={2011},
   month=aug }

@article{Chavanis::2016,
   title={Collapse of a self-gravitating Bose-Einstein condensate with attractive self-interaction},
   volume={94},
   ISSN={2470-0029},
   url={http://dx.doi.org/10.1103/PhysRevD.94.083007},
   DOI={10.1103/physrevd.94.083007},
   number={8},
   journal={Physical Review D},
   publisher={American Physical Society (APS)},
   author={Chavanis, Pierre-Henri},
   year={2016},
   month=oct }

\appendix

\begin{widetext}

\section{SIGNIFICANT QUANTITIES OF THE NEWTONIAN GAUGE}
\label{App-A}

\textbf{CHRISTOFFEL SYMBOL}

\begin{gather}
    \tensor{\Gamma}{^\eta_\eta_\eta} = \; \frac{1}{a} \frac{d a}{d \eta} + \frac{ \partial_\eta \Phi}{1 + 2 \, \Phi}\,, \quad
    \tensor{\Gamma}{^r_r_r} = - \frac{\partial_r \Psi}{1 - 2 \, \Psi}\,, \quad
    \tensor{\Gamma}{^\eta_\eta_r} = \; \frac{\partial_r \Phi}{1 + 2 \, \Phi }\,, \quad 
    \tensor{\Gamma}{^\eta_\eta_\theta} = \; \frac{\partial_\theta \Phi}{1 + 2 \, \Phi }\,, \quad 
    \tensor{\Gamma}{^\eta_\eta_\varphi} = \; \frac{\partial_\varphi \Phi}{1 + 2 \, \Phi }\,, \nonumber \\
    \tensor{\Gamma}{^r_\eta_\eta} = \; \frac{\partial_r \Phi}{1 - 2 \, \Psi}\,, \quad
    \tensor{\Gamma}{^\theta_\eta_\eta} = \; \frac{\partial_\theta \Phi}{r^{2} \left(1 - 2 \, \Psi \right)}\,, \quad
    \tensor{\Gamma}{^\varphi_\eta_\eta} = \; \frac{\partial_\varphi \Phi}{{r^{2} \sin^2 (\theta) \left(1 - 2 \, \Psi\right)}}\,, \nonumber \\
    \tensor{\Gamma}{^\eta_r_r} = \; \frac{\tensor{\Gamma}{^\eta_\theta_\theta}}{r^{2}} = \frac{\tensor{\Gamma}{^\eta_\varphi_\varphi}}{r^{2} \sin^2 (\theta)} = \frac{\frac{1}{a} \frac{d a}{d \eta} {\left(1 - 2 \, \Psi \right)} - \partial_\eta \Psi}{1 + 2 \, \Phi}\,, \quad
    \tensor{\Gamma}{^r_\eta_r} = \; \tensor{\Gamma}{^\theta_\eta_\theta} = \tensor{\Gamma}{^\varphi_\eta_\varphi}  = \frac{1}{a} \frac{d a}{d \eta} - \frac{\partial_\eta \Psi}{1 - 2 \, \Psi}\,, \nonumber \\
    \tensor{\Gamma}{^\theta_r_\theta} = \; \tensor{\Gamma}{^\varphi_r_\varphi} = - \frac{\tensor{\Gamma}{^r_\theta_\theta}}{r^{2}} = - \frac{\tensor{\Gamma}{^r_\varphi_\varphi}}{r^{2} \sin^2 (\theta)} = \frac{1}{r} - \frac{\partial_r \Psi}{1 - 2 \, \Psi}\,, \quad
    \tensor{\Gamma}{^\theta_\theta_\theta} = \tensor{\Gamma}{^r_r_\theta} = - r^{2} \; \tensor{\Gamma}{^\theta_r_r} =  - \frac{\partial_\theta \Psi}{1 - 2 \, \Psi}\,, \nonumber \\
    \tensor{\Gamma}{^\varphi_\varphi_\varphi} = \tensor{\Gamma}{^r_r_\varphi} = \tensor{\Gamma}{^\theta_\theta_\varphi} = - \sin^2 (\theta) \; \tensor{\Gamma}{^\varphi_\theta_\theta} = - r^{2} \sin^2 (\theta) \; \tensor{\Gamma}{^\varphi_r_r} = -\frac{\partial_\varphi \Psi}{1 - 2 \, \Psi} \,, \nonumber \\
    \tensor{\Gamma}{^\theta_\varphi_\varphi} = -\sin^2 (\theta)\; \tensor{\Gamma}{^\varphi_\theta_\varphi} = -\sin(\theta) \cos(\theta) + \frac{\sin^2 (\theta) \partial_\theta \Psi}{1 - 2 \, \Psi}\,, \nonumber \\
    \tensor{\Gamma}{^\eta_r_\theta} \, = \,  \tensor{\Gamma}{^\eta_r_\varphi} \, = \, \tensor{\Gamma}{^\eta_\theta_\varphi} \, = \, \tensor{\Gamma}{^r_\eta_\theta} \, = \, \tensor{\Gamma}{^r_\eta_\varphi} \, = \, \tensor{\Gamma}{^r_\theta_\varphi} = \tensor{\Gamma}{^\theta_\eta_r} \, = \, \tensor{\Gamma}{^\theta_\eta_\varphi} \, = \, \tensor{\Gamma}{^\theta_r_\varphi} \, = \, \tensor{\Gamma}{^\varphi_\eta_\theta} \, = \, \tensor{\Gamma}{^\varphi_\eta_r} \, = \, \tensor{\Gamma}{^\varphi_r_\theta} \, = \,0\,.
\end{gather}

\textbf{EINSTEIN TENSOR}

\begin{align}
    \tensor{G}{_\eta_\eta} =& \; \frac{3}{a^2} \left(\frac{d a}{d \eta}\right)^{2} - \frac{6}{a} \frac{d a}{d \eta} \frac{\partial_\eta \Psi}{1 - 2\, \Psi} + \frac{3 \left(\partial_\eta \Psi \right)^2}{\left(1 - 2\, \Psi \right)^2} + \frac{2 \left(1 + 2 \, \Phi \right)}{\left(1 - 2\, \Psi \right)^2} \nabla^2 \Psi + \frac{3 \left(1 + 2 \, \Phi \right)}{\left(1 - 2\, \Psi \right)^3} \left(\nabla \Psi\right)^{2} \,, \\
    \tensor{G}{_\eta_r} =& \; \frac{2}{a} \frac{d a}{d \eta} \frac{\partial_r \Phi}{1 + 2\, \Phi} + \frac{2 \, \partial_\eta \partial_r \Psi}{1 - 2\, \Psi} - \frac{2 \, \partial_\eta \Psi \, \partial_r \Phi}{\left(1 + 2\, \Phi \right) \left(1 - 2 \, \Psi \right)} + \frac{4 \, \partial_\eta \Psi \, \partial_r \Psi}{\left(1 - 2 \, \Psi \right)^2} \,, \\
    \tensor{G}{_\eta_\theta} =& \; \frac{2}{a} \frac{d a}{d \eta} \frac{\partial_\theta \Phi}{1 + 2\, \Phi} + \frac{2 \, \partial_\eta \partial_\theta \Psi}{1 - 2\, \Psi} - \frac{2 \, \partial_\eta \Psi \, \partial_\theta \Phi}{\left(1 + 2\, \Phi \right) \left(1 - 2 \, \Psi \right)} + \frac{4 \, \partial_\eta \Psi \, \partial_\theta \Psi}{\left(1 - 2 \, \Psi \right)^2} \,, \\
    \tensor{G}{_\eta_\varphi} =& \; \frac{2}{a} \frac{d a}{d \eta} \frac{\partial_\varphi \Phi}{1 + 2\, \Phi} + \frac{2 \, \partial_\eta \partial_\varphi \Psi}{1 - 2\, \Psi} - \frac{2 \, \partial_\eta \Psi \, \partial_\varphi \Phi}{\left(1 + 2\, \Phi \right) \left(1 - 2 \, \Psi \right)} + \frac{4 \, \partial_\eta \Psi \, \partial_\varphi \Psi}{\left(1 - 2 \, \Psi \right)^2} \,, \\
    \tensor{G}{_r_r} =& - \frac{2 \left(1 - 2\, \Psi \right)}{1 + 2\, \Phi} \frac{1}{a} \frac{d^2 a}{d \eta^2} + \frac{1 - 2 \, \Psi}{1 + 2\, \Phi} \frac{1}{a^2} \left(\frac{d a}{d \eta}\right)^{2} + \frac{2 \left(1 - 2\, \Psi \right)}{\left(1 + 2\, \Phi \right)^2} \frac{1}{a} \frac{d a}{d \eta} \partial_\eta \Phi + \frac{4}{1 + 2\, \Phi} \frac{1}{a} \frac{d a}{d \eta} \partial_\eta \Psi + \frac{2 \, \partial_\eta^2 \Psi}{1 + 2\, \Phi} \nonumber \\
    &+ \frac{\left(\partial_\eta \Psi\right)^{2}}{\left(1 + 2\, \Phi \right) \left(1 - 2 \, \Psi\right)} - \frac{2 \, \partial_\eta \Phi \, \partial_\eta \Psi}{\left(1 + 2\, \Phi \right)^2} + \frac{\nabla^2 \Phi - \partial_r^2 \Phi}{1 + 2 \, \Phi} - \frac{\nabla^2 \Psi - \partial_r^2 \Psi}{1 - 2\, \Psi} - \frac{\left(\nabla \Phi\right)^{2} - \left(\partial_r \Phi\right)^{2}}{\left(1 + 2\, \Phi \right)^2} \nonumber \\
    &- \frac{2 \left(\nabla \Psi\right)^{2} - 3 \left(\partial_r \Psi\right)^{2}}{\left(1 - 2\, \Psi \right)^2} - \frac{2 \, \partial_r \Phi \, \partial_r \Psi}{\left(1 + 2\, \Phi \right) \left(1 - 2 \, \Psi\right)} \, , \\
    \tensor{G}{_r_\theta} =& - \frac{r}{1 + 2\, \Phi} \, \partial_r \left( \frac{\partial_\theta \Phi}{r} \right) + \frac{r}{1 - 2\, \Psi} \, \partial_r \left( \frac{\partial_\theta \Psi}{r} \right)  + \frac{\partial_r \Phi \, \partial_\theta \Phi}{\left(1 + 2\, \Phi \right)^2} + \frac{3 \, \partial_r \Psi \, \partial_\theta \Psi}{\left(1 - 2\, \Psi \right)^2} - \frac{\partial_\theta \Phi \, \partial_r \Psi + \partial_r \Phi \, \partial_\theta \Psi}{\left(1 + 2\, \Phi \right) \left(1 - 2 \, \Psi \right)} \,, \\
    \tensor{G}{_r_\varphi} =& - \frac{r}{1 + 2\, \Phi} \, \partial_r \left( \frac{\partial_\varphi \Phi}{r} \right) + \frac{r}{1 - 2\, \Psi} \, \partial_r \left( \frac{\partial_\varphi \Psi}{r} \right) + \frac{\partial_r \Phi \, \partial_\varphi \Phi}{\left(1 + 2\, \Phi \right)^2} + \frac{3 \, \partial_r \Psi \, \partial_\varphi \Psi}{\left(1 - 2\, \Psi \right)^2} - \frac{\partial_\varphi \Phi \, \partial_r \Psi + \partial_r \Phi \, \partial_\varphi \Psi}{\left(1 + 2\, \Phi \right) \left(1 - 2 \, \Psi\right)} \,, \\
    \frac{\tensor{G}{_\theta_\theta}}{r^2} =& - \frac{2 \left(1 - 2\, \Psi \right)}{1 + 2\, \Phi} \frac{1}{a} \frac{d^2 a}{d \eta^2} + \frac{1 - 2\, \Psi}{1 + 2\, \Phi} \frac{1}{a^2} \left(\frac{d a}{d \eta}\right)^{2} + \frac{2 \left(1 - 2\, \Psi \right)}{\left(1 + 2\, \Phi \right)^2} \frac{1}{a} \frac{d a}{d \eta} \partial_\eta \Phi + \frac{4}{1 + 2\, \Phi} \frac{1}{a} \frac{d a}{d \eta} \partial_\eta \Psi + \frac{2 \, \partial_\eta^2 \Psi}{1 + 2\, \Phi} \nonumber \\
    &+ \frac{\left( \partial_\eta \Psi \right)^2}{\left(1 + 2\, \Phi \right) \left(1 - 2 \, \Psi\right)} - \frac{2 \, \partial_\eta \Phi \, \partial_\eta \Psi}{\left(1 + 2\, \Phi \right)^2} + \frac{\nabla^2 \Phi - \frac{1}{r} \partial_r \Phi - \frac{1}{r^2} \partial_\theta^2 \Phi}{1 + 2\, \Phi} - \frac{\nabla^2 \Psi - \frac{1}{r} \partial_r \Psi - \frac{1}{r^2} \partial_\theta^2 \Psi}{1 - 2\, \Psi} \nonumber \\
    &- \frac{\left(\nabla \Phi\right)^{2} - \frac{1}{r^2} \left(\partial_\theta \Phi\right)^{2}}{\left(1 + 2\, \Phi \right)^2} - \frac{2 \left(\nabla \Psi\right)^{2}  - \frac{3}{r^2} \left(\partial_\theta \Psi\right)^{2}}{\left(1 - 2\, \Psi \right)^2} - \frac{\frac{2}{r^2} \, \partial_\theta \Phi \, \partial_\theta \Psi}{\left(1 + 2\, \Phi \right) \left(1 - 2 \, \Psi\right)} \,, \\
    \tensor{G}{_\theta_\varphi} =& - \frac{\sin(\theta)}{1 + 2\, \Phi} \, \partial_\theta \left( \frac{\partial_\varphi \Phi}{\sin(\theta)} \right) + \frac{\sin(\theta)}{1 - 2\, \Psi} \, \partial_\theta \left( \frac{\partial_\varphi \Psi}{\sin(\theta)} \right) + \frac{\partial_\theta \Phi \, \partial_\varphi \Phi}{\left(1 + 2\, \Phi \right)^2} + \frac{3 \, \partial_\theta \Psi \, \partial_\varphi \Psi}{\left(1 - 2\, \Psi \right)^2} \nonumber \\
    &- \frac{\partial_\theta \Phi \, \partial_\varphi \Psi + \partial_\varphi \Phi \, \partial_\theta \Psi}{\left(1 + 2\, \Phi \right) \left(1 - 2 \,\Psi \right)} \, , \\
    \frac{\tensor{G}{_\varphi_\varphi}}{r^2 \sin^2 (\theta)} =& - \frac{2 \left(1 - 2\, \Psi \right)}{1 + 2\, \Phi} \frac{1}{a} \frac{d^2 a}{d \eta^2} + \frac{1 - 2\, \Psi}{1 + 2\, \Phi} \frac{1}{a^2} \left(\frac{d a}{d \eta}\right)^{2} + \frac{2 \left(1 - 2\, \Psi\right)}{\left(1 + 2\, \Phi\right)^2} \frac{1}{a} \frac{d a}{d \eta} \partial_\eta \Phi + \frac{4}{1 + 2\, \Phi} \frac{1}{a} \frac{d a}{d \eta} \partial_\eta \Psi + \frac{2 \, \partial_\eta^2 \Psi}{1 + 2\, \Phi} \nonumber \\
    &+ \frac{\left(\partial_\eta \Psi\right)^{2}}{\left(1 + 2\, \Phi \right) \left(1 - 2 \, \Psi\right)} - \frac{2 \, \partial_\eta \Phi \, \partial_\eta \Psi}{\left(1 + 2\, \Phi \right)^2} + \frac{\partial_r^2 \Phi + \frac{1}{r} \partial_r \Phi + \frac{1}{r^2} \partial_\theta^2 \Phi}{1 + 2\, \Phi} - \frac{\partial_r^2 \Psi + \frac{1}{r} \partial_r \Psi + \frac{1}{r^2} \partial_\theta^2 \Psi}{1 - 2\, \Psi} \nonumber \\
    &- \frac{\left(\nabla \Phi\right)^2 - \frac{1}{r^2 \sin^2 (\theta)} \left(\partial_\varphi \Phi\right)^2}{\left(1 + 2\, \Phi \right)^2} - \frac{2 \left(\nabla \Psi\right)^2 - \frac{3}{r^2 \sin^2 (\theta)} \left(\partial_\varphi \Psi\right)^2}{\left(1 - 2\, \Psi \right)^2} - \frac{\frac{2}{r^2 \sin^2(\theta)} \partial_\varphi \Phi \, \partial_\varphi \Psi}{\left(1 + 2\, \Phi \right) \left(1 - 2 \, \Psi\right)} \,.
\end{align}

\textbf{CURVATURE SCALAR}

\begin{align}
    R =& \; \frac{6}{a^3 \left(1 + 2\, \Phi \right)}\frac{d^2 a}{d\eta^2} - \frac{6 \, \partial_\eta \Phi}{a^3 \left(1 + 2\, \Phi \right)^2} \frac{da}{d \eta} - \frac{18 \, \partial_\eta \Psi}{a^3 \left(1 + 2 \, \Phi \right) \left(1 - 2 \, \Psi \right)} \frac{da}{d \eta} + \frac{6 \, \partial_\eta \Phi \, \partial_\eta \Psi}{a^2 \left(1 + 2\, \Phi \right)^2 \left(1 - 2\, \Psi\right)} \nonumber \\
    &- \frac{6 \, \partial_\eta^2 \Psi}{a^2 \left(1 + 2\, \Phi \right) \left(1 - 2\, \Psi\right)} -\frac{2 \, \nabla^2 \Phi}{a^2 \left(1 + 2\, \Phi \right) \left(1 - 2\, \Psi\right)} + \frac{4 \, \nabla^2 \Psi}{a^2 \left(1 - 2\, \Psi\right)^2} + \frac{2 \left( \nabla \Phi \right)^2}{a^2 \left(1 + 2\, \Phi \right)^2 \left(1 - 2\, \Psi\right)} \nonumber \\
    &+ \frac{6 \left( \nabla \Psi \right)^2}{a^2 \left(1 - 2\, \Psi \right)^3} + \frac{2 \, \nabla \Phi \cdot \nabla \Psi}{a^2 \left(1 + 2\, \Phi \right) \left(1 - 2\, \Psi\right)^2} \,.
\end{align}

\textbf{DOUBLE COVARIANT DERIVATIVE ON A SCALAR}

\begin{align}
    \nabla_\eta \nabla_\eta \, \phi =& \; \partial_\eta^2 \phi - \frac{1}{a} \frac{d a}{d \eta} \partial_\eta \phi - \frac{ \partial_\eta \Phi \, \partial_\eta \phi}{1 + 2 \, \Phi} - \frac{\nabla \Phi \cdot \nabla \phi}{1 - 2 \, \Psi} \,,\\
    \nabla_\eta \nabla_r \, \phi =& \; \partial_\eta \partial_r \phi - \frac{1}{a} \frac{d a}{d \eta} \partial_r \phi - \frac{\partial_r \Phi \, \partial_\eta \phi}{1 + 2\, \Phi} + \frac{\partial_\eta \Psi \, \partial_r \phi}{1 - 2\, \Psi} \,, \\
    \nabla_\eta \nabla_\theta \, \phi =& \; \partial_\eta \partial_\theta \phi - \frac{1}{a} \frac{d a}{d \eta} \partial_\theta \phi - \frac{\partial_\theta \Phi \, \partial_\eta \phi}{1 + 2\, \Phi} + \frac{\partial_\eta \Psi \, \partial_\theta \phi}{1 - 2\, \Psi} \,,\\
    \nabla_\eta \nabla_\varphi \, \phi =& \; \partial_\eta \partial_\varphi \phi - \frac{1}{a} \frac{d a}{d \eta} \partial_\varphi \phi - \frac{\partial_\varphi \Phi \, \partial_\eta \phi}{1 + 2\, \Phi} + \frac{\partial_\eta \Psi \, \partial_\varphi \phi}{1 - 2\, \Psi} \,,\\
    \nabla_r \nabla_r \, \phi =& \; \partial_r^2 \phi - \frac{1}{a} \frac{d a}{d \eta} \frac{1 - 2 \, \Psi}{1 + 2 \, \Phi} \partial_\eta \phi +\frac{\partial_\eta \Psi \, \partial_\eta \phi}{1 + 2 \, \Phi} - \frac{\nabla \Psi \cdot \nabla \phi}{1 - 2 \, \Psi} + \frac{2 \, \partial_r \Psi \, \partial_r \phi}{1 - 2 \, \Psi} \,,\\
    \nabla_r \nabla_\theta \, \phi =& \; r \, \partial_r \left( \frac{\partial_\theta \phi}{r} \right) + \frac{\partial_\theta \Psi \partial_r \phi + \partial_r \Psi \, \partial_\theta \phi}{1 - 2\, \Psi} \,,\\
    \nabla_r \nabla_\varphi \, \phi =& \; r \, \partial_r \left(\frac{\partial_\varphi \phi}{r} \right) + \frac{\partial_\varphi \Psi \, \partial_r \phi + \partial_r \Psi \, \partial_\varphi \phi}{1 - 2\, \Psi} \,,\\
    \frac{\nabla_\theta \nabla_\theta \, \phi}{r^2} =& \; \frac{\partial_\theta^2 \phi}{r^2} - \frac{1}{a} \frac{d a}{d \eta} \frac{1 - 2 \, \Psi}{1 + 2 \, \Phi} \partial_\eta \phi + \frac{\partial_\eta \Psi \, \partial_\eta \phi}{1 + 2 \, \Phi} + \frac{\partial_r \phi}{r} - \frac{\nabla \Psi \cdot \nabla \phi}{1 - 2 \, \Psi} + \frac{2 \, \partial_\theta \Psi \, \partial_\theta \phi}{r^2 \left(1 - 2 \, \Psi\right)} \,,\\
    \nabla_\theta \nabla_\varphi \, \phi =& \; \sin (\theta) \, \partial_\theta \left( \frac{\partial_\varphi \phi}{\sin (\theta)} \right) + \frac{\partial_\varphi \Psi \, \partial_\theta \phi + \partial_\theta \Psi \, \partial_\varphi \phi}{1 - 2\, \Psi} \,,\\
    \frac{\nabla_\varphi \nabla_\varphi \, \phi}{r^2 \sin^2 (\theta)} =& \; \frac{\partial_\varphi^2 \phi}{r^2 \sin^2 (\theta)} - \frac{1}{a} \frac{d a}{d \eta} \frac{1 - 2 \, \Psi}{1 + 2 \, \Phi} \partial_\eta \phi + \frac{\partial_\eta \Psi \, \partial_\eta \phi}{1 + 2 \, \Phi} + \frac{\partial_r \phi}{r} + \frac{\cos (\theta)}{r^2 \sin (\theta)} \partial_\theta \phi - \frac{\nabla \Psi \cdot \nabla \phi}{1 - 2\, \Psi} \nonumber \\
    &+ \frac{2 \, \partial_\varphi \Psi \, \partial_\varphi \phi}{r^2 \sin^2(\theta) \left(1 - 2\, \Psi\right)} \,.
\end{align}

\textbf{THE COVARIANT D'ALEMBERTIAN ON AN SCALAR}

\begin{align}
    \nabla^\alpha \, \nabla_\alpha \, \phi =& - \frac{\partial_\eta^2 \phi}{a^2 \left(1 + 2\, \Phi \right)} + \frac{\nabla^2 \phi}{a^2 \left(1 - 2\, \Psi \right)} - \frac{2}{a^3} \frac{d a}{d \eta} \frac{\partial_\eta \phi}{1 + 2\, \Phi} + \frac{ \partial_\eta \Phi \, \partial_\eta \phi}{a^2 \left(1 + 2 \, \Phi\right)^2} + \frac{3\, \partial_\eta \Psi \, \partial_\eta \phi}{a^2 \left(1 + 2 \, \Phi\right) \left(1 - 2 \, \Psi\right)} \nonumber \\
    &+ \frac{\nabla \Phi \cdot \nabla \phi}{a^2 \left(1 + 2 \, \Phi\right) \left(1 - 2 \, \Psi\right)} - \frac{\nabla \Psi \cdot \nabla \phi}{a^2 \left(1 - 2 \, \Psi\right)^2} \,.
\end{align}

\end{widetext}
\end{document}